\documentclass[conference]{IEEEtran}
\DeclareUnicodeCharacter{202F}{~}
\IEEEoverridecommandlockouts
\usepackage{cite}
\usepackage{amsmath,amssymb,amsfonts}
\usepackage{algorithmic}
\usepackage{graphicx}
\usepackage{textcomp}
\usepackage{xcolor}
\usepackage[utf8]{inputenc}
\usepackage{pifont}
\usepackage{tabularx}
\usepackage{amssymb}
\usepackage{pifont}
\usepackage{array} 
\usepackage{tabularx}
\usepackage{makecell}
\newcommand{\xmark}{\ding{55}}
\renewcommand{\checkmark}{\ding{51}}
\def\BibTeX{{\rm B\kern-.05em{\sc i\kern-.025em b}\kern-.08em
    T\kern-.1667em\lower.7ex\hbox{E}\kern-.125emX}}
\begin{document}

\title{Design Process of a Self Adaptive Smart Serious Games Ecosystem \\ 

\thanks{*Research supported by Chinese Scholarship Council, (202308390066, 202508390057) \\
X. Tao (xiya.t@alumnos.upm.es), P. Chen (peng.chen@alumnos.upm.es), M. Tsami  (mariaeleni.tsami@alumnos.upm.es), F. Khayati (fedi.khayati@alumnos.upm.es), and
M. Eckert (corresponding author phone: 00 34 91 067 87 62, e-mail: martina.eckert@upm.es) are with the Research Center on Software Technologies and Multimedia Systems for Sustainability (CITSEM), Universidad Politécnica de Madrid (UPM), Spain. 
}
}

\author{\IEEEauthorblockN{X. Tao, P. Chen, M. Tsami, F. Khayati, M. Eckert}
}

\maketitle

\begin{abstract}
This paper outlines the design vision and planned evolution of Blexer v3, a modular and AI-driven rehabilitation ecosystem based on serious games. Building on insights from previous versions of the system, we propose a new architecture that aims to integrate multimodal sensing, real-time reasoning, and intelligent control. The envisioned system will include distinct modules for data collection, user state inference, and gameplay adaptation. Key features such as dynamic difficulty adjustment (DDA) and procedural content generation (PCG) are also considered to support personalized interventions. We present the complete conceptual framework of Blexer v3, which defines the modular structure and data flow of the system. This serves as the foundation for the next phase: the development of a functional prototype and its integration into clinical rehabilitation scenarios. 

\end{abstract}


\section{Introduction}
Video games have evolved significantly since their inception in the 1960s, becoming a cultural force in the late 1980s and early 1990s \cite{DON_2010_ReplayHistoryVideo}. With the growth of the videogame industry, games have expanded into fields such as education, military, and healthcare, known as Serious Games (SGs) \cite{ABT_1970_SeriousGames}. In healthcare, SGs have shown promise in screening \cite{ZYG_2020_DetectionMildCognitivea} and rehabilitation \cite{MAR_2020_PhysiolandSeriousGamea}. These games use physiological and behavioral data from wearable sensors in real-time to deliver engaging experiences that promote adherence and recovery. 

In particular, the integration of Artificial Intelligence (AI) into SGs has shown potential to personalize therapeutic content and improve engagement \cite{DAM_2023_SeriousGamesGamification}. Techniques such as neural networks (NN) and reinforcement learning (RL) assess user states and adjust game difficulty in real time, allowing personalized interventions aligned with therapeutic goals \cite{TAO_2025_MappingLandscapeArtificial}.

Despite recent progress, most SG-based rehabilitation systems are tightly coupled to specific games and therapeutic scenarios. These closed-loop designs often lack modularity and generalization, hindering cross-context adaptation. There is a clear need for flexible, multi-game systems that provide personalized rehabilitation strategies. 

Since 2015, we have explored adaptive systems for rehabilitation-focused SGs, although efforts were limited by AI and sensing capabilities. Recent advances make real-time personalisation feasible. The use of AI in SGs remains fragmented and often game-specific \cite{TAO_2025_MappingLandscapeArtificial}.

To address this, we propose a generalizable framework that decouples user modeling and intelligent control from individual games. Our system approach is still in the design process and currently proposes the use of a convolutional neural network (CNN) together with a recurrent neural network (RNN) to process multimodal physiological signals \cite{PAT_2025_IntegrationSeriousGamesa,GON_2024_DeepLearningApproacha}, in addition to a Large Language Model  (LLM) to provide the possibility of dynamic dialogues and narrative adaptation. Explainable AI tools (e.g. LIME, SHAP) will be used to support transparency and trust \cite{RIB_2016_WhyShouldTrusta}, and a shared AI core will enable scalable, human-centred interaction in multiple rehabilitation games \cite{GON_2024_DeepLearningApproacha}.The novelty of Blexer v3 lies not only in the design of individual modules, but particularly in their specific combination: CAM as a centralized reasoning core and IPM as a universal actuator, enabling adaptive control across multiple games.

The article is structured as follows: after a short introduction to Dynamic Difficulty Adjustment (DDA) in section \ref{DDA}, which is applied to ensure sustained engagement and therapeutic effectiveness, we present the historic development of the Blexer System in section \ref{Background}, which forms the basis for our proposal of a new modular architecture for SG ecosystems. Then, section \ref{Blexer_enh} explains the proposal in detail and \ref{conclusion} conludes the article.

\section{DDA}\label{DDA}
Dynamic Difficulty Adjustment was originally developed to maintain a player in a flow state \cite{GUO_2024_RethinkingDynamicDifficulty}. to assure adherance and motivation. Now, it is also a widely used mechanism to personalize the gameplay of SGs to optimize the outcome through better and longer play. Given that static difficulty settings are inadequate for accommodating individual variations in performance, emotion, or physiological states\cite{ANG_2017_ComparingEffectsDynamic}, DDA can dynamically modify challenge levels by analyzing player behavior and internal feedback in real time, making it particularly useful for our system.

\subsection{Foundations and Motivation}
The achievement of a suitable balance between challenge and skill, the maintenance of user engagement and the maintenance of smooth transitions without disrupting immersion \cite{GUO_2024_RethinkingDynamicDifficulty} can be supported by the integration of gameplay-derived indicators and psychophysiological measures. Gameplay-based approaches are based on behavioral indicators such as error rates and completion time to infer player performance and engagement patterns \cite{KOR_2023_VisualizationVirtualReality}. In contrast, psychophysiological methods extract information from biometric data (e.g. heart rate) and emotional signals (e.g. facial expressions) to estimate physical load and affective state \cite{LAR_2021_InductionEmotionalStates}.
Our system incorporates both types of input to enhance accuracy and responsiveness to adaptation to difficulties. Specifically, we have implemented modules for heart rate monitoring and facial emotion recognition, enabling real-time inference of internal states that are not directly observable from gameplay alone. This multimodal sensing approach allows our DDA mechanism to respond more precisely to user needs, especially in rehabilitation scenarios where emotional engagement and fatigue are crucial. Further technical details will be presented in section \ref{Blexer_enh}.

\subsection{Classification and Implementation}
In addition to using gameplay and physiological data, the DDA needs to define a policy to determine how the system should respond to these inputs and adjust the difficulty of the game in real time\cite{MOR_2024_DynamicDifficultyAdjustment}.

DDA policies are commonly classified into two main categories: rule-based approaches and data-driven approaches. Rule-based systems rely on manually designed rules that define how difficulty should be modified under specific conditions, often using thresholds or logical mappings. The strategies for determining such thresholds have been discussed in our previous work \cite{ECK_2023_FindingEffectiveAdjustment}. These methods are straightforward, easy to interpret, and well suited to controlled environments such as rehabilitation games. Data-driven approaches, on the contrary, aim to learn adaptive policies from empirical data, allowing greater flexibility, generalization, and personalization. The choice between the two typically depends on the availability of training data, the desired adaptability, and the system transparency requirements.

Although rule-based systems offer robustness and transparency, they are limited in expressiveness and scalability\cite{GUO_2024_RethinkingDynamicDifficulty}. In response, many recent studies have explored data-driven approaches, including classical machine learning models that learn difficulty adjustment policies from annotated player behavior data \cite{XUE_2017_DynamicDifficultyAdjustment}. Some systems also apply clustering or regression techniques to capture user preferences and performance dynamics. Although more flexible, these approaches are often sensitive to data quality and require careful feature design, limiting their adaptability in real-time applications.

More recently, neural network-based approaches have been employed to model complex, nonlinear relationships between player state and game difficulty. Among these, reinforcement learning (RL) has emerged as a promising strategy to optimize DDA policies through direct interaction with the game environment. RL-based systems dynamically adjust difficulty to maximize long-term outcomes such as engagement or emotional balance, and have shown particular potential in rehabilitation contexts \cite{KHA_2024_MasteringSuperMario, HUB_2021_DynamicDifficultyAdjustment}. However, these approaches require carefully designed reward structures and substantial computational resources.

\subsection{Current state of art of difficulty adaptation}
Recent research has advanced the use of data-driven methods, particularly RL, to develop more responsive and autonomous DDA systems. These approaches allow games to learn adaptation strategies based on long-term interactions with the player, using feedback signals such as engagement, task success, or physiological changes to guide decision making \cite{ZHE_2024_DynamicDifficultyAdjustment}. RL-based DDA is especially suited to scenarios that require gradual and personalized progression, such as serious games for rehabilitation or cognitive training \cite{RAH_2023_ContinuousReinforcementLearningbased, ZHA_2024_ImplementationEffectEvaluation}. 

As these systems become more adaptive, attention has turned to not only adjusting difficulty but also dynamically shaping the content of the game. Procedural Content Generation (PCG) provides a means of generating levels, tasks, and scenarios that respond to inferred player states. When combined with DDA mechanisms, PCG can continuously tailor both the challenge and structure of the game experience. This integration has been applied in contexts such as virtual reality games and adaptive training tasks, where maintaining engagement between repeated sessions is critical \cite{HUB_2021_DynamicDifficultyAdjustment}.

These combined techniques lay the foundation for the adaptive logic that will be used in our system. We propose a modular system where two modules interact to achieve the DDA: the central Context-Aware Module (CAM) knows the state of the user and their current needs, takes the decisions accordingly, and communicates with the Intelligent Play Module (IPM), which then executes the difficulty adaptation and sends back the user's performance as a feedback to the CAM. However, as a starting point, our current implementation adopts a rule-based strategy in which difficulty levels are determined by predefined conditional logic based on the previous results \cite{ECK_2023_FindingEffectiveAdjustment}. In this way, we create a test game that later will be included in the whole architecture and process as detailed in Section \ref{Blexer_enh}.

\section{Background}\label{Background}
The Blexer system, developed by the research Group for Acoustics and Multimedia Applications (GAMMA), Universidad Politécnica de Madrid (UPM), is a telerehabilitation platform designed for patients with physical disabilities. Since its inception in 2016, the system has evolved through multiple versions, transitioning from static configurations to fully adaptive intelligent interventions. The following summarizes the technical evolution across its versions.

\subsection{Blexer v1: Web-supervised rehabilitation with fixed parameters}
Blexer v1 combined Kinect-based motion capture (Kinect Xbox 360) with games developed in Blender and used a middleware called Chiro to connect to a web-based platform, Blexer-med \cite{ECK_2022_NewArchitectureCustomizable}. Therapists could configure training parameters-such as goals and time limits-for each patient via the platform. These configurations were downloaded to the patient’s PC in JavaScript Object Notation (JSON) format and performance results were uploaded after each session, allowing remote supervision and manual reconfiguration.

The system implemented the idea of full-scale exergames, surpassing the conventional mini-game structures. The exemplar game, Phiby's Adventures, features four training activities—wood cutting, tree climbing, rowing, and diving—embedded within a purpose-driven adventure map. Through motion amplification, patients with restricted mobility could perform complete in-game actions, thereby boosting the sense of immersion. This system has been evaluated numerous times across various settings, including individuals with diverse neuromuscular disorders \cite{ECK_2022_NewArchitectureCustomizable}, children with cerebral palsy \cite{ECK_2023_FindingEffectiveAdjustment}, and more recently, elderly stroke survivors \cite{PEL_2023_UseVirtualReality}.

Although Blexer v1 supported web-based configuration and asynchronous feedback, it lacked real-time adaptability and could not sense the player’s emotional or physiological states.

\begin{figure}
    \centering
    \includegraphics[width=1\linewidth]{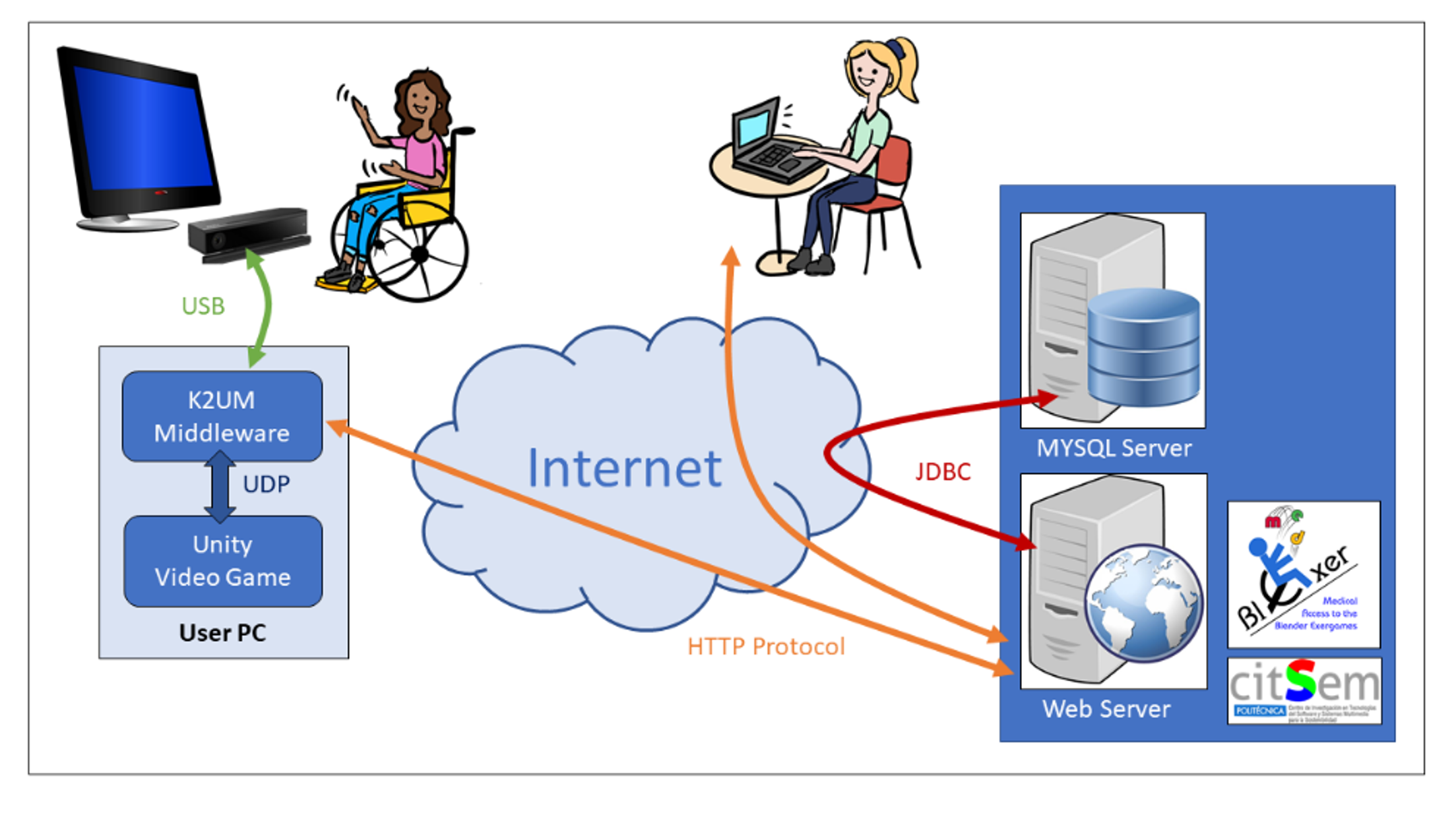}
    \caption{Architecture of the Blexer ecosystem version 2}
    \label{fig:Blexer_v2-framework}
\end{figure}

\subsection{Blexer v2: Performance-based personalisation via manual adjustments}
Blexer v2 has progressed to using the Kinect Xbox One and integrates the games within Unity, maintaining its original framework as illustrated in Fig. \ref{fig:Blexer_v2-framework}. The middleware, now referred to as Kinect to Unity Middleware (K2UM), transfers motion capture data from the Kinect sensor to the games and interfaces with the Blexer-med web platform. This platform is employed by therapists to modify the difficulty settings for each exercise (target quantity and time limit) in each game tailored to individual patients. 
To aid decision-making, graphical feedback was integrated into the Blexer-med user interface to help therapists visualise performance trends and make faster and more informed adjustments. The rules for these decisions are based on the results obtained in \cite{ECK_2023_FindingEffectiveAdjustment}.

However, Blexer v2 still presents several challenges:
\begin{itemize}
    \item The modifications rely on the observations of the therapist, subjective judgments, and manual parameter tuning. This has the drawback that therapists often lack time, do not monitor patients during exercises, and may make incorrect choices due to stress or an incomplete understanding of the patient's present physical or emotional condition or other circumstances.
    \item The system cannot respond to changes in real-time in fatigue, disengagement, or emotional changes.
    \item The system is limited to processing motor data, which narrows the scope for making decisions and identifying changes in the user that warrant alerts.
\end{itemize}

Therefore, although Blexer v2 allows for greater customization in comparison to v1, it still functions as a system guided by human input with restricted autonomous intelligence. This limitation has driven the creation of the more advanced Blexer v3. This new iteration is presently in development, yet we aim to move forward with the novel framework structure and the latest additions in sensors and affect estimation.

\section{Current work on Blexer v3: Context-aware and AI-driven adaptation}\label{Blexer_enh}

\begin{figure}
    \centering
    \includegraphics[width=1\linewidth]{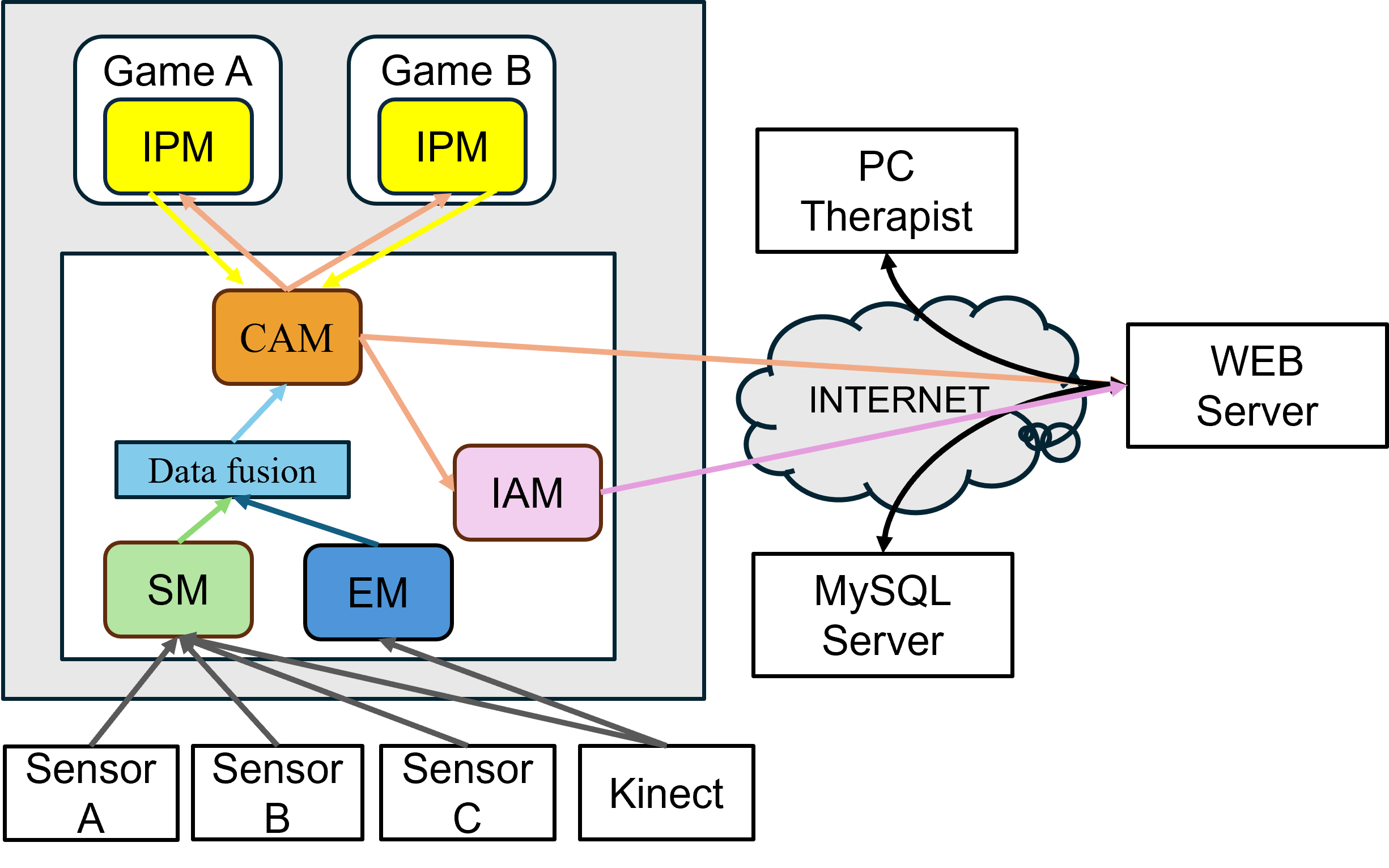}
    \caption{Modules added to middleware and games to compound the new architecture of the Blexer ecosystem version 3}
    \label{fig:Blexerv3_modules}
\end{figure}

To overcome the limitations of conventional therapist-driven rehabilitation systems, the third generation of the Blexer framework will be based on a modular and intelligent architecture designed for context-sensitive adaptation in real time. The platform is being developed to integrate multiple sensing modalities, including motion, emotion, and physiological data, along with AI-driven modules for decision making and real-time feedback. As illustrated in Fig. \ref{fig:Blexerv3_modules}, the architecture of Blexer v3 will probably include:

\begin{itemize}
    \item \textbf{Sensor Module (SM):} Support of various sensors, including Kinect, custom cameras, and additional devices (e.g., heart rate sensors, pulse monitors), providing vital physiological data.
    
    \item \textbf{Emotion Module (EM):} Facial expression analysis for emotion detection, using the RGB image provided by the MoCap camera.
    
    \item \textbf{Context Awareness Module (CAM):} Real-time context analysis considering sensor data, player mood, game progress, past game history, medical recommendations, player preferences, time of day and other factors that could be relevant in decision making.
    
    \item \textbf{Intelligent Play Module (IPM):} Intelligent game control using CAM data. Adaptation of the gameplay to suit the patient’s needs, ensuring they follow therapist instructions while avoiding excessive fatigue or discouragement.
    
    \item \textbf{Interactive Assistance Module (IAM):} Support for therapists by continuously monitoring patient performance, providing real-time feedback, and notifying in the event of abnormalities, for example, increase in fatigue.

 \end{itemize}  

\subsection{Sensor module}\label{sensor_module_description}

The Blexer system sensor module (SM) has already been implemented and is responsible for the real-time acquisition, processing, and synchronization of multimodal physiological and behavioral data. It acts as the primary interface between the user’s physical state and the logic of the game, transforming raw sensor inputs into structured, time-aligned information streams that support personalized interventions and adaptive gameplay. 

The objective behind the SM's design is to broaden sensing coverage by incorporating not just motion data, but also biometric signals from wearable sensors, thus facilitating precise, real-time physiological monitoring.

The module collects data from heterogeneous sources, including skeletal posture captured by a Mocap camera (Kinect), as well as physiological signals such as heart rate acquired via Bluetooth Low Energy (BLE) devices. We selected the Polar H10 chest strap, a device that has undergone extensive testing and validation in various studies.\cite{SCH_2022_ValidityPolarH10}. In addition, we sought an appropriate wristband or smartwatch for ease of use by the player, as these options are simpler to wear and remove and less prone to rejection. Regrettably, we couldn't find any commercial device that allows for direct connection and raw data access due to protective measures; most systems offer data only through their own applications. This could complicate integration into our system and introduce undesirable latency. Ultimately, we successfully incorporated the Bangle.js open source programmable smartwatch\cite{Espruino}, which provides reasonably reliable data; it is not as precise as the Polar H10, but still meets our requirements.

With these two devices, our system currently supports the following physiological parameters:

\begin{itemize}
    \item \textbf{Heart Rate (BPM)}: Instantaneous heart rate estimates from both Polar H10 and Bangle.js 2, displayed in the middleware dashboard in real time.
    \item \textbf{RR Intervals (IBI)}: High resolution Polar H10 RR intervals, converted to milliseconds for analysis of heart rate variability.
    \item \textbf{Accelerometer Data}: Raw X, Y, Z values from Bangle.js 2, scaled to represent gravitational units (Gs).
    \item \textbf{Measurement Confidence}: A Bangle.js 2 signal quality score is reported along with heart rate data.
\end{itemize}

All incoming data undergo local preprocessing-including noise filtering, format standardization, and timestamp alignment-and are subsequently transmitted as unified JSON packets through the middleware to the Unity game engine, while also being stored for real-time use and offline analysis. 

\begin{figure}
  \centering
  \includegraphics[width=1\linewidth]{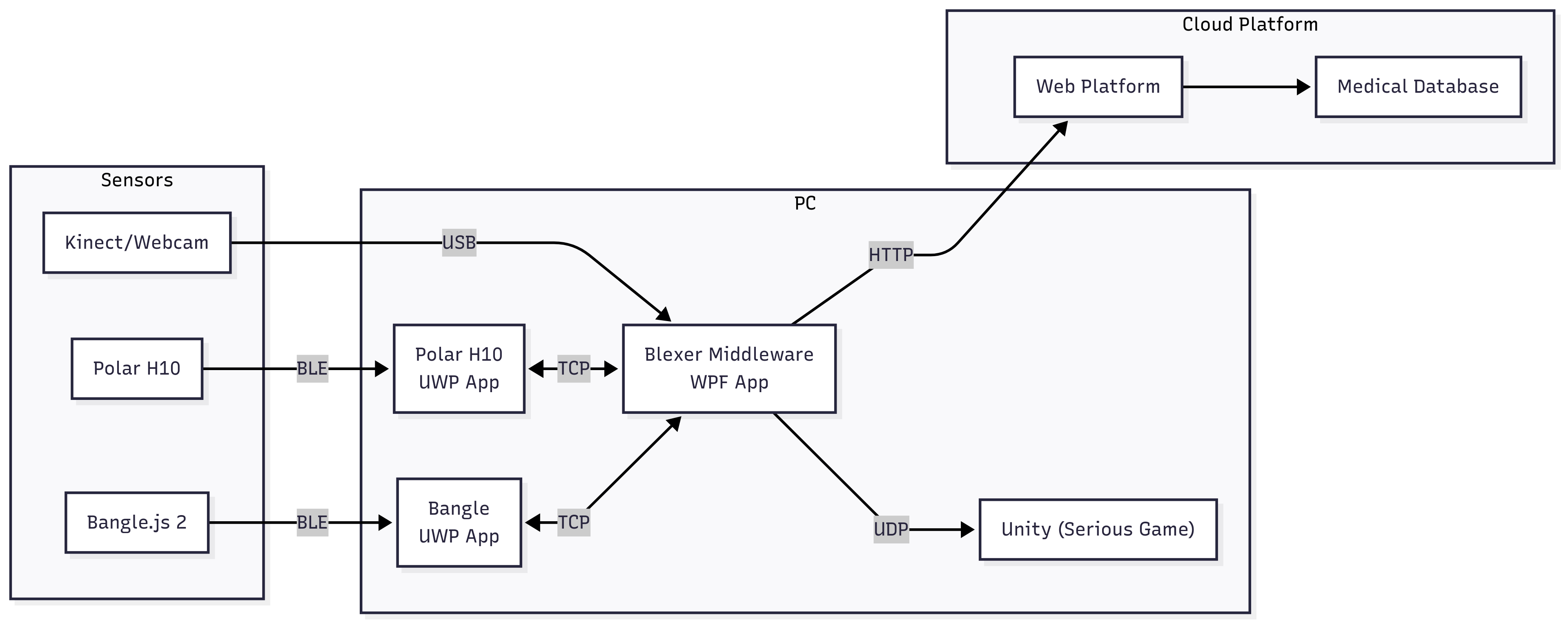}
  \caption{System Architecture}
  \label{fig:architecture}
\end{figure}

\subsubsection{\textbf{Heart Rate Monitoring}}
As a key component of the physiological sensing pipeline, the heart rate monitoring subsystem provides continuous access to the internal states of users through the wearables. It accommodates both the \textbf{electrocardiography (ECG)} and \textbf{photoplethysmography (PPG)} methods, providing a compromise between signal accuracy and ease of use.

For ECG-based acquisition, high-precision heart rate (BPM) and RR interval (IBI) data are captured and parsed by a custom Universal Windows Platform (UWP) application. The processed data are formatted into structured strings and transmitted via Transmission Control Protocol (TCP) to the middleware \ref{fig:architecture}. Each notification includes the current BPM and a sequence of RR intervals with ±1 ms, converted from 1/1024 s units to milliseconds for analysis and visualization of heart rate variability (HRV).

In parallel, the PPG-based channel provides BPM estimates, three-axis accelerometer data (X, Y, Z), and signal confidence scores via the Nordic UART BLE protocol. These streams are received by a local edge device (e.g. laptop or Raspberry Pi), time-stamped, and structured into JSON packets before being relayed to the middleware over TCP for integration with other modalities.

Although the current implementation focuses on ECG and PPG data streams, the architecture supports extensibility. In particular, gyroscope data could be integrated to capture fine wrist movements, such as rotations, not detectable by camera-based motion tracking.

For development and debugging purposes, the middleware includes a visualization panel (“Sensor Graphs”) to display BPM, RR intervals, and motion data in real time, as well as a diagnostics tab to monitor TCP communications and data packet integrity. Robust handshake and disconnect protocols are implemented to ensure stable operation on multiple BLE devices.

\subsubsection{\textbf{Emotion Module}}

To enable real-time affective feedback during the game, we added the Emotion Module (EM) based on facial emotion recognition (FER). This component performs real-time emotion inference locally and streams skeletal and affective data in structured JSON format over the User Datagram Protocol (UDP) for integration into the Context Awareness Module (CAM). 

The FER sub-module integrates two main computational stages: face detection and emotion classification. For face detection, we adopted a Deep Neural Network (DNN) architecture based on SSD with a ResNet-10 backbone, accessed via OpenCV’s DNN module. This model processes RGB frames captured by the Kinect sensors, extracting the most prominent of the detected faces. The face region is then passed to a CNN trained on the FER-2013 data set\cite{fer2013} for emotion prediction. The data set contains facial images collected under varying head poses and lighting conditions, which makes it suitable for robust emotion recognition in real-world game scenarios.

\begin{figure}
    \centering
    \includegraphics[width=1\linewidth]{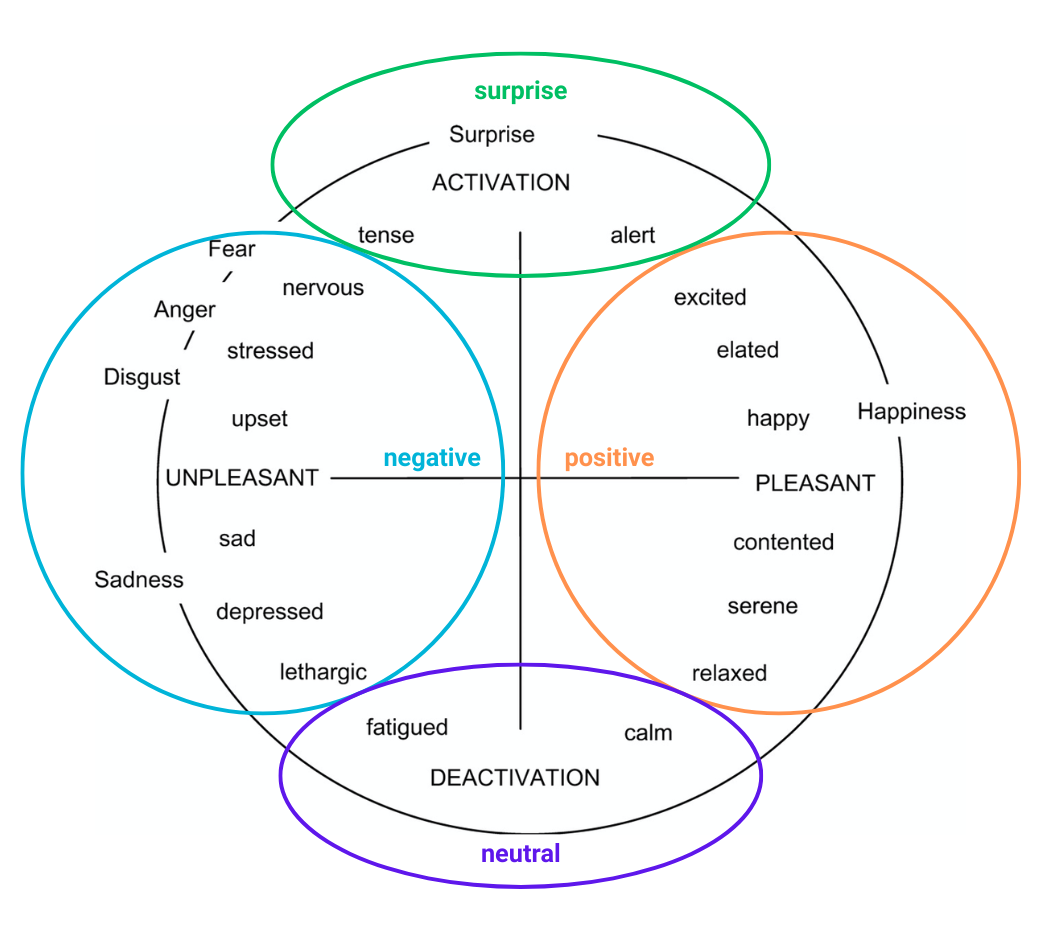}
    \caption{Emotion classes adopted in our FER model, visualization on the  Russell circumflex model \cite{Russell}}
    \label{fig:Russ}
\end{figure}

To enhance robustness and real-time performance, the original seven-label emotion set (anger, disgust, fear, happiness, sadness, surprise, neutral) was reduced to four core states: positive, neutral, surprise, and negative (in Fig. \ref{fig:Russ}). Based on Russell’s circumplex model \cite{Russell}, closely clustered emotions such as fear and disgust were grouped under negative to improve reliability in unconstrained gameplay. This simplified structure better supports modules like DDA and engagement analysis by focusing on general affective trends rather than fine-grained distinctions.

The CNN consists of four convolutional blocks (64–512 filters) with batch normalization, ReLU, max pooling, and dropout, followed by an L2-regularized fully connected layer. Training applied data augmentation (rotation, zoom) and callbacks (EarlyStopping, ReduceLROnPlateau). Although results remain suboptimal, future work will build a dataset of “gaming expressions” (e.g., concentration, tension) that fall outside conventional emotion categories.

The model, developed in TensorFlow/Keras, was converted to ONNX for .NET Framework 4.8 compatibility and integrated into the Blexer middleware, enabling synchronized real-time FER feedback in Unity.

\subsection{\textbf{Multimodal Data Fusion}}

Multimodal data fusion is a central architectural principle in Blexer v3, intended to support a coherent and context-sensitive understanding of the user. Rather than treating raw sensor streams in isolation, the system adopts a layered fusion strategy that gradually transforms heterogeneous input into structured, semantically rich representations. This transformation enables downstream reasoning modules to make informed, personalized, and clinically grounded decisions.

The first level of fusion occurs within the Sensor Module. As described previously, this module synchronizes and pre-processes various physiological and behavioral signals, including heart rate, RR intervals, skeletal tracking, and facial expressions, into a consistent temporal format. Each modality undergoes filtering, normalization, and time-stamp alignment. In the case of facial emotion data, the raw output of the FER model is reduced to a small set of affective states (positive, neutral, surprise, and negative) to enhance interpretability and real-time performance. This low-latency data layer provides an integrated snapshot of the user’s physical and emotional state, and serves as the foundational input for higher level modules.

The second level of fusion occurs within the Context Awareness Module (CAM), where the focus shifts from signal harmonization to multimodal interpretation. Here, pre-fused sensor data is combined with gameplay records, therapist-defined goals, user preferences, and contextual information (e.g., time of day, session history). CAM then applies mid-level fusion mechanisms, such as cross-modal attention, to infer latent user states such as cognitive workload, emotional engagement, and physical fatigue. These inferred states form the basis for adaptive decision making, which guides gameplay adjustments through the Intelligent Play Module (IPM).

In this layered scheme, data fusion acts as a semantic bridge between sensing and decision. It ensures that raw signals are not only cleaned and synchronized, but also meaningfully interpreted in relation to game context and therapeutic intent. By embedding fusion across both the Sensor Module and CAM, Blexer v3 is designed to maintain both responsiveness and semantic depth, ultimately allowing more adaptive, explainable and personalized rehabilitation experiences.

\subsection{\textbf{Context Awareness Module (CAM)}}

The Context Awareness Module (CAM) is designed to serve as the centralized reasoning and decision-making core within the modular architecture of Blexer v3. Unlike game-specific controllers, CAM is intended to operate independently of any single title, by integrating large-scale multimodal data streams--including physiological, emotional, behavioral, and contextual information--into a unified representation of the user’s state. This broader scope allows CAM to make clinically relevant high-level decisions that can be transferred across different rehabilitation games.

Rather than handling raw signals directly, CAM builds upon the semantically enriched inputs produced by the Sensor Module and the Data Fusion component. These inputs reflect the physiological and emotional state of the user in context, providing a coherent basis for high-level inference and adaptive decision making. 

\begin{figure}
    \centering
    \includegraphics[width=1\linewidth]{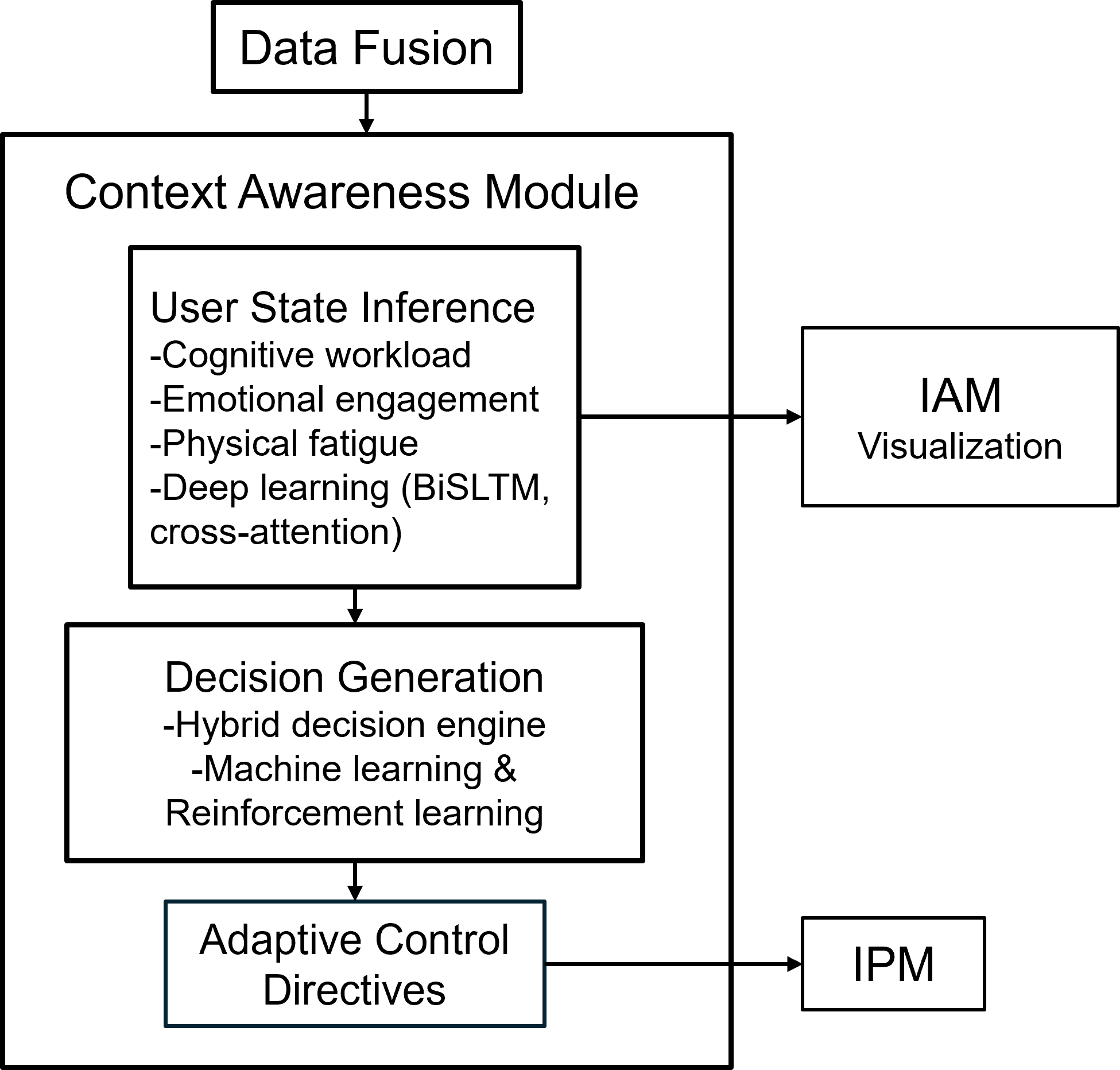}
    \caption{Data flow of the Context Awareness Module (CAM) }
    \label{fig:Russ2}
\end{figure}

As illustrated in Fig.~\ref{fig:Russ2}, CAM’s reasoning pipeline is structured into two stages. The first stage, \textbf{user state inference}, is planned to estimate latent variables such as cognitive workload, emotional engagement, and physical fatigue. These inferences are derived from patterns across multiple input streams, rather than from any isolated signal. For example, a combination of elevated heart rate, deterioration of motion smoothness, and flat facial affect can be interpreted as a sign of task-related stress or deactivation. To support this process, CAM is designed to take advantage of deep learning techniques, including temporal modeling with BiLSTM, which helps capture sequential changes in user state, and mid-level multimodal fusion methods such as cross-attention, which allow the system to integrate the most informative features from different modalities. These AI-based methods enable CAM to capture both short-term fluctuations and longer-term trends in the user state with improved accuracy and adaptability.

The state estimates obtained in the first stage are treated as intermediate representations rather than final system outputs. Their role is to support the second stage, \textbf{decision generation}, where CAM is planned to determine appropriate adaptation strategies based on the current user profile. To do so, CAM will employ a hybrid decision engine composed of both rule-based and learning-based components. Specifically, the learning-based component is intended to incorporate machine learning and reinforcement learning algorithms, enabling CAM to refine its strategies over time based on gameplay feedback and therapeutic outcomes. The rule-based engine will ensure safe and predictable behavior under clinical constraints, such as automatically reducing difficulty when physical fatigue exceeds a predefined threshold.

In the second stage, CAM is designed to generate \textbf{adaptive control directives} that guide how the game environment should respond to the user’s condition. Rather than selecting a specific minigame, CAM will recommend a suitable \textbf{type or category of task}, such as coordination, reaction speed, or memory-based activities. These recommendations are expected to include personalized gameplay parameters, such as the number of required repetitions, expected task duration, pacing suggestions, or target difficulty levels. CAM may also indicate the need for rest intervals or a reduced intensity of stimulus when fatigue or low engagement is inferred.

All outputs from CAM are intended to be formatted as structured, standardized commands and passed to the IPM, which acts as the execution agent within the system. CAM does not alter gameplay mechanics directly; instead, it provides general adaptation strategies that IPM interprets in a game-agnostic manner. To enhance this interaction, a dedicated instruction format is planned for future versions, allowing CAM directives, such as task categories, repetition counts, or pacing adjustments, to be expressed in a more consistent and interoperable way. This separation of reasoning and execution ensures that CAM remains focused on high-level adaptation, while IPM handles game-level implementation in varying therapeutic contexts.

In addition, CAM is designed to assist therapists and support \textbf{transparency} through its integration with the \textbf{Insight and Administration Module (IAM)}, which serves as the system’s monitoring and reporting interface. Real-time visualizations of inferred user states and generated control strategies will be available via IAM dashboards. This functionality is particularly valuable for remote or asynchronous rehabilitation, where therapists may not be physically present but still require timely insights into user performance and system behavior. Such visual summaries can help clinicians adjust therapy plans, evaluate engagement levels, and ensure safety over extended home use.

CAM is also planned to be extensible. Its architecture is designed to support integration with future inference models, external clinical data sources, or user-specific historical profiles. For example, a future AI-driven version may incorporate long-term therapy data to adjust adaptation logic based on patient progress trajectories or comorbidities. These developments would further increase the personalization and effectiveness of the rehabilitation experience.

In summary, CAM is designed to bridge low-level multimodal sensing and high-level gameplay adaptation. By interpreting physiological, emotional, behavioral, and contextual data, it is intended to produce user-tailored, clinically relevant directives that support meaningful participation and therapeutic progress. While its core logic and architectural structure have been defined, full implementation, integration, and validation will take place in the next stages of system development.

\subsection{\textbf{Intelligent Play Module (IPM)}}

The Intelligent Play Module (IPM) functions as a universal actuator that interprets CAM-generated directives and applies them in real time across different rehabilitation games. Unlike a purely game-agnostic execution layer, the IPM is aware of the underlying game logic and mechanics, enabling it to translate high-level adaptation strategies into concrete in-game adjustments. In this way, IPM can modify gameplay parameters, feedback intensity, and content structure in a manner that preserves consistency with CAM’s reasoning while remaining sensitive to the specific requirements of each game.

\begin{itemize}
    \item Task difficulty (e.g., number of repetitions, time constraints);
    \item Exercise switching in response to detected fatigue or boredom;
    \item Intensity of feedback and duration of the session.
\end{itemize}
In addition, the IPM incorporates a PCG method to support the dynamic creation and reordering of exercise content and game levels, enhancing engagement and variability throughout the training process.

the Intelligent Play Module (IPM) receives structured adaptive instructions from the CAM mentioned previously and applies them in real-time to modify game content and interaction mechanics. Working as actuator within the game environment, IPM ensures that the gameplay dynamically aligns with both the therapeutic guidelines and the evolving physiological and emotional state of the player. therefore, this module primarily plays the excution role in our system.

\subsubsection{Personalized In-Game Adaptation}

IPM implements a set of generic control mechanisms capable of interpreting CAM-generated instructions in a game-agnostic manner. Based on real-time input from CAM-including cognitive workload, emotional engagement, and physical fatigue-IPM dynamically adjusts gameplay parameters such as:

\begin{itemize}
\item \textbf{Exercise Selection and Scheduling:} Prioritizing or rearranging tasks to reflect player preferences and performance trends.
\item \textbf{Interaction Feedback:} Adjusting visual, auditory, or haptic responses to maintain user engagement while avoiding overstimulation.
\end{itemize}

Following CAM's context-driven recommendations, IPM ensures that players are continually challenged within their optimal developmental zone\cite{GUO_2024_RethinkingDynamicDifficulty}, while avoiding frustration or fatigue by the DDA
policy introduced in Section~\ref{DDA}.

\subsubsection{Communication between CAM and IPM}
A central novelty of Blexer v3 lies in the explicit design of the interaction between the Context Awareness Module (CAM) and the Intelligent Play Module (IPM). Unlike prior rehabilitation systems where adaptation logic was either embedded directly in the game or handled through ad hoc rules, our architecture introduces a \emph{dedicated reasoning--execution division}. CAM operates as the high-level reasoning core, generating adaptive directives from multimodal user data, while a single system-level IPM functions as the universal actuator, translating these directives into game-specific actions across multiple rehabilitation titles. 

This explicit CAM--IPM interface is a new design element that ensures both \emph{consistency} of adaptation strategies and \emph{flexibility} for heterogeneous games. By decoupling reasoning from execution, the architecture provides scalability (one IPM that coordinates many exercises), transparency (directives expressed in structured JSON schemas), and clinical safety (centralized enforcement of thresholds and fallback strategies). This separation of concerns represents a shift from game-tied adaptation toward a reusable control framework at the ecosystem-level.

\subsubsection{Our prototype}

An initial prototype of the IPM has been integrated into an adventure-style rehabilitation game that includes four foundational exercise scenarios: alternating arm lifts, arm raise and hammering motion, forward–backward arm movements, and left–right body shifts. At this stage, the system operates under fixed rules and predefined action sequences, allowing for controlled testing of the system architecture and its response to adaptive feedback. These exercises serve as a functional testbed for validating core interaction mechanisms prior to the integration of intelligent control. 

Within this environment, a rule-based IPM is designed to receive instructions regarding in-game task adjustments, such as modifying repetition counts, task duration, and sequence ordering. For instance, when the CAM detects signs of physical fatigue during an alternating arm task, it can recommend a reduction in repetition frequency or substitution with a more engaging task that maintains therapeutic value. The functional responsibilities of CAM and IPM in such adaptive processes are summarized in Table~\ref{tab:cam_ipm_functions}. 

\begin{table}[htbp]
\caption{Functional Division Between CAM and IPM}
\label{tab:cam_ipm_functions}
\centering
\scriptsize
\renewcommand{\arraystretch}{1.2}
\begin{tabularx}{\linewidth}{|>{\raggedright\arraybackslash}X|
                             >{\raggedright\arraybackslash}X|
                             >{\raggedright\arraybackslash}X|}

\hline
\textbf{Function Category} & \textbf{Belongs to CAM} & \textbf{Belongs to IPM} \\
\hline
Reduce repetition frequency & \checkmark\ Yes (state-based inference) & \xmark\ No \\
\hline
Feedback intensity adjustment  & \checkmark\ Yes (strategic selection) & \checkmark\ Yes (feedback execution) \\
\hline
Modify specific in-game parameter  & \xmark\ No & \checkmark\ Yes (in-game behavior) \\
\hline
Select game category & \checkmark\ Yes & \xmark\ No \\
\hline
Select specific mini-game ID & \xmark\ No & \checkmark\ Yes (IPM or game logic) \\
\hline
\end{tabularx}
\end{table}

This modular architecture offers clear separation of responsibilities: CAM functions as a centralized reasoning unit for state interpretation and strategy generation, while IPM serves as the execution layer for local game adaptation. This decoupling enhances system flexibility, reusability, and scalability. Future developments of the IPM are planned to incorporate advanced DDA and PCG techniques, enabling not only real-time personalization of challenge levels but also adaptive control over game structure and content.

\section{Conclusion \label{conclusion}}
The development of Blexer v3 will hopefully represent a significant step toward intelligent and personalized rehabilitation through serious games. By integrating multimodal real-time sensing, intelligent algorithms, and modular architecture, the system will allow for continuous adaptation of the game based on the cognitive, emotional, and physiological states of the user. This could improve therapeutic outcomes and reduce clinician workload by offering automated support and transparent feedback.

The modular separation between context inference (CAM) and execution (IPM) allows the system to scale across various rehabilitation scenarios and game types. In addition, the integration of Dynamic Difficulty Adjustment (DDA) with Procedural Content Generation (PCG) opens the door to continuous, engaging, and individualized rehabilitation experiences.

In the near future, a limited study will be conducted with a reduced group of mean players to test the feasibility and detect necessary improvements. Later, more advanced reinforcement learning models will be tried to improve personalization. In general, Blexer v3 is expected to demonstrate the feasibility of emotionally responsive and data-driven therapeutic ecosystems.

% References

\bibliographystyle{IEEEtran}
\bibliography{references}

\end{document}